\documentclass[usenatbib]{mn2e}
\usepackage{epsfig}
\input epsf

\begin{document}

\title[Spectroscopy of a Globular Cluster in NGC 6822]{Spectroscopy of a Globular Cluster in the Local Group dIrr NGC 6822}

\author[Jay Strader et al.]{
Jay Strader,$^{1}$\thanks{Electronic mail: strader@ucolick.org}
Jean P. Brodie,$^{1}$
John P. Huchra$^{2}$
\\
$^1$\,UCO/Lick Observatory, University of California, Santa Cruz, CA 95064 \\
$^2$\,Harvard-Smithsonian Center for Astrophysics, Cambridge, MA 02138 \\
}

\date{Accepted ... Received ...}

\maketitle

\begin{abstract}

We present low-resolution Keck spectroscopy for the globular cluster (GC) H VIII in the Local Group dIrr galaxy NGC 6822. We find the metallicity of the
cluster to be [Fe/H]=$-1.58\pm0.28$ and the age of the cluster to be 3--4 Gyr, slightly older than but consistent with previous age estimates. H VIII
seems to be more metal-poor than most intermediate-age GCs in the Local Group, and appears most similar to the anomalous Small Magellanic Cloud (SMC)
clusters Lindsay 113 and NGC 339.

\end{abstract}


\section{Introduction}

Local Group dwarf irregular (dIrr) galaxies such as the Magellanic Clouds (MCs) seem to possess an interesting distinction compared to other Local Group
galaxies in that they contain intermediate-age (1 Gyr $\la$ age $\la 10$ Gyr) globular clusters (GCs; where we define GCs as rich, centrally concentrated
star clusters, with no age connotation). By contrast, both the giant spirals M31 and the Milky Way (MW) and the dwarf spheroidal (dSph) galaxies, while
often having star formation histories as complex as the dIrrs, appear to contain GC populations that are almost uniformly old ($\sim 10-13$ Gyr). However,
there may be a few exceptions: some metal-rich M31 clusters and Terzan 7 in the Sagittarius dSph appear to be $\sim 8-9$ Gyr old (Barmby \& Huchra 2000;
Buonanno et al. 1995).

The unique GC systems of dIrrs suggest that these gas-rich galaxies offer a good environment for studying recent/ongoing GC formation. Of the $\sim 16$ dIrr
galaxies in the Local Group, four are known to have GCs: the MCs, NGC 6822, and the Wolf-Lundmark-Melotte (WLM) galaxy. The last of these has only one GC,
which is old ($14.8\pm0.6$ Gyr; Hodge et al. 1999). Despite their populous systems of star clusters, the MCs are not an ideal laboratory for studying dwarf
galaxy evolution, due to the close (and likely influential) proximity of the MW. The more isolated NGC 6822 offers a better ``closed-box'' environment for
study, and is similar enough to the MCs to allow direct comparison of cluster properties. In this paper, we analyze Keck spectra of two GC candidates in
order to measure metallicities, ages, and abundances for the clusters.

\section{Observations and Data Reduction}

Two GC candidates, H IV and H VIII, were identified from Hubble's original candidate list \citep{H25}, as studied in \citet{H77}. They were observed in
May 2001 using both the blue and red sides of LRIS (Low Resolution Imaging Spectrograph; Oke et al. 1995) in longslit mode on the Keck I telescope. A single
600s exposure was taken for each of the $V \sim 17-18$ clusters. On the blue side, we used a 400 l/mm grism, which gave a resolution of $\sim 8$ \AA\ and a
dispersion of 1.74 \AA/pix. On the red side, a 600 l/mm grating, blazed at 5000 \AA, was utilized. This provided a resolution of $\sim 6$ \AA\ and a
dispersion of 1.28 \AA/pix. The useable wavelength range of the blue spectra was about 3700 -- 5300 \AA, and that of the red spectra 5800 -- 8000 \AA.

The standard data reduction was performed using IRAF\footnote{IRAF is distributed by the National Optical Astronomy Observatories, which is operated
by the Association of Research in Astronomy, Inc., under cooperative agreement with the National Science Foundation, U.S.A.}. Raw images were debiased and
then flatfielded, using a normalized composite sky flat. After cosmic ray removal using the IRAF task \emph{cosmicrays}, the spectra were wavelength
calibrated with HgNeArCdZn comparison arc lamp spectra. These spectra were then flux-calibrated using the flux standard PG 1708+602.

The wavelength calibration for the blue spectra was challenging due to the paucity of arc lamp calibration lines in the blue, particularly in the vicinity of
H$\beta$. The result of this was that after the ``standard" calibration was completed, the poor extrapolation of the polynomial solution to longer wavelengths
resulted in a shift of ten \AA\ or higher in H$\beta$ from its laboratory value. To partially alleviate this problem, after making a minor radial velocity
correction (using the systemic velocity of NGC 6822, $-57\pm2$ km/s;  de Vaucouleurs et al. 1993) we fixed the wavelength solution at H$\delta$ (4101~\AA) and
H$\beta$ (4861~\AA). This allowed accurate measurements of all of the indices of interest, which generally lay within a few hundred angstroms of those two
points.

While the spectrum of H VIII is that of a star cluster (see Fig.~1 for the blue spectrum), it is clear from the spectrum of H IV that it is an
H\,{\sevensize\bf II} region, as previously suggested by spectroscopy \citep[hereafter C00]{C00} and \emph{Hubble Space Telescope (HST)} photometry 
\citep[hereafter W00]{W00}. Thus, our subsequent analysis focuses solely on the bona fide GC H VIII. 

\section{Results}

\subsection{Metallicity}

\citet[hereafter BH90]{BH90} defined a procedure for estimating the metallicity of globular clusters with low resolution spectra by taking a weighted mean of
various elemental absorption-line indices sensitive to [Fe/H]. Indices defined over a wavelength range of interest are calculated with respect to a
pseudocontinuum, defined using regions to either side of the feature bandpass.

We derived a value of [Fe/H] = $-1.58\pm0.28$ for H VIII using the BH90 formulae, with a few modifications. No resolution correction was made for 
measuring BH90 metallicity indices, since the BH90 calibration primarily used data near our resolution or better. The indices Mg2, MgH, and Fe5270 had part
of their continuum bandpasses truncated by the long wavelength limit of the blue spectrum (5300 \AA). Thus, we excluded these three indices in estimating the
total metallicity. The Ca H+K index was also excluded, since the Ca H line blended with the strong Balmer line H$\epsilon$. We also calculated the variance of
the final [Fe/H] estimate slightly differently than BH90, as described in \citet{LB02}. The individual index [Fe/H] estimates (as well as the composite value)
are listed in Table 1. It should be kept in mind that since the BH90 method was calibrated using old globular clusters, it may be less accurate for measuring
the metallicity of younger stellar populations.

Due to the low S/N and poor wavelength coverage of their H VIII spectrum, C00 were unable to determine a metallicity for this cluster. For comparison, other
clusters in NGC 6822 have been found to be similarly metal-poor: C00 estimate [Fe/H] = $-1.46\pm0.26$ for the very young ($70\pm10$ Myr) cluster H VI. This is
more metal-rich than the old ($\sim 11$ Gyr) GC H VII, for which \citet{CB98} find [Fe/H] = $-1.95\pm0.15$ and C00 give [Fe/H] $\sim -2.0$.

\subsection{Lick/IDS Indices and Cluster Ages}

Elemental abundances were measured using Lick/IDS absorption-line indices (Trager et al.~1998 and references therein), from spectra convolved with a 5-pixel
(equivalent to 6.4 \AA) Gaussian kernel to more closely match the Lick/IDS resolution. See \citet{LB02} and \citet{S02} for more extensive discussions of
the resolution matching issue.

We first measured a set of Lick/IDS indices (Table 2) and Balmer line indices (Table 3; some of these are defined on the Lick/IDS system) for 
comparison to stellar evolutionary models. As discussed earlier, some of the most commonly-used indices (e.g., Mg2, Mgb, Fe5270, Fe5355), which
had part of their continuum or feature bandpasses truncated by the red limit of the blue spectrum, could not be measured. To estimate the age
of H VIII, we plotted [Fe/H] vs. Balmer line diagrams from \citet{MT00} models, using the [Fe/H] estimate obtained from the BH90 procedure. An
[Fe/H] vs. H$\beta$ is given in Figure 2, and an [Fe/H] vs. H$\gamma_{A}$ plot in Figure 3.

The first of these figures suggests that H VIII is $\sim 5^{+ 2}_{- 1}$ Gyr old. The second plot gives a younger age of $3.5\pm1$ Gyr. There may
be cause to give more weight to the H$\gamma_{A}$ estimate, as that index is generally less metal dependent than H$\beta$, and there is a $\sim 0.3$ dex
uncertainty in our estimate of the cluster [Fe/H]. (The downside of using H$\gamma$ (or H$\delta$) is that since the index is more narrow than H$\beta$,
higher S/N spectra are needed to obtain accurate measurements). In sum, our data favor an age of 3--4 Gyr for H VIII.

W00 found an age of $1.8\pm0.2$ Gyr by fitting isochrones to their \emph{HST} color-magnitude diagram (CMD). However, the recommended best-fit isochone does
not seem a particularly good match to the data, although older isochrones were not included. Moreover, the CMD suffers from heavy field contamination (see
their Fig.~7), suggesting they are underestimating their errors. Our age estimate is consistent with that of C00, $1.4^{+ 2.2}_{- 0.5}$ Gyr, calculated using
integrated colors and the equivalent width of H$\beta$. Together, these studies indicate H VIII is $\sim 3\pm1$ Gyr old and we suggest that, with realistic
errors, the W00 data are also consistent with this result.

\section{Discussion}

These derived parameters for H VIII place it in a relatively unpopulated area of age-metallicity phase space for Local Group GCs. While there are a
significant number of intermediate-age GCs in the LMC ($\sim 15$; Bica et al. 1998), these all have [Fe/H] $\ga -1.0$. The intermediate-age GCs
in the SMC have a much wider range in metallicities (Mighell et al. 1998; Da Costa \& Hatzidimitriou 1998, hereafter DH98), from [Fe/H] $\sim -0.2$ to $-1.7$
(DH98; Piatti et al. 2001). The luminosity of H VIII is M$_{V} \sim -5.6$, lower than that of most of the SMC clusters.

H VIII may have its closest analogues in two clusters studied by DH98: Lindsay 113 (L113) and NGC 339 (N339). Using the $W\arcmin$ Ca {\sevensize\bf 
II} triplet method, they find that L113 and N339 have [Fe/H] = $-1.44\pm0.16$ and $-1.46\pm0.10$, respectively. The age of L113 is $6\pm1$ Gyr \citep{S87},
and that of N339 is $4 - 6$ Gyr (DH98; Mighell et al. 1998), although \citet{R00} argue N339 is $\sim 8$ Gyr old, which would make it more consistent with
the SMC cluster age-metallicity relation. They find the most likely explanation for these anomalous clusters to be an infall of less-enriched gas $\sim 5-6$
Gyr ago, either from the Magellanic Stream, or more speculatively, from a bound, early epoch gas clump.

It is more difficult to tell whether H VIII is an anomalous cluster in its environment, considering the paucity of the NGC 6822 cluster population. The
presence of cluster H VI (which has essentially the same [Fe/H], see \S 3.1) indicates that little chemical enrichment has taken place in the last several 
Gyr, in contrast to the LMC and SMC. This might then indicate that the continuing enrichment in the MCs is a function of their tidal interactions with the
MW, and that ``normal'' isolated dIrrs have a relatively constant star formation rate (SFR), possibly punctuated by stochastic increases which result in star
cluster formation. \citet{G96} report a doubling of the SFR over the long-term mean in NGC 6822 over the last few hundred Myr (see discussion in Cohen \&
Blakeslee 1998), and it seems possible that H VI formed due to this stochastic SFR enhancement, and that H VIII may have had a similar origin several Gyr
ago.

\section{Conclusions}

We have observed two GC candidates in the Local Group dIrr NGC 6822 with LRIS on the Keck I telescope. We found that H IV is an H {\sevensize\bf II}
region and that H VIII is a GC. We derived [Fe/H]=$-1.58\pm0.28$ for H VIII. We have estimated the age of the cluster to be 3--4 Gyr, slightly older
than but consistent with previous age estimates. It seems to be more metal-poor than most intermediate-age GCs in the Local Group, and appears most similar
to the anomalous Small Magellanic Cloud clusters Lindsay 113 and NGC 339. We have briefly considered the chemical evolution of NGC 6822, and suggest that H
VIII and the young cluster H VI may have formed during stochastic peaks of the global star formation rate, which overall seems to have been relatively
constant during the lifetime of the galaxy.

\section*{Acknowledgments}

We thank Doug Geisler and an anonymous referee for helpful comments. This work was supported by NSF grants AST-9900732 and AST-0206139.

\begin{center}
\begin{figure*}   
\psfig{file=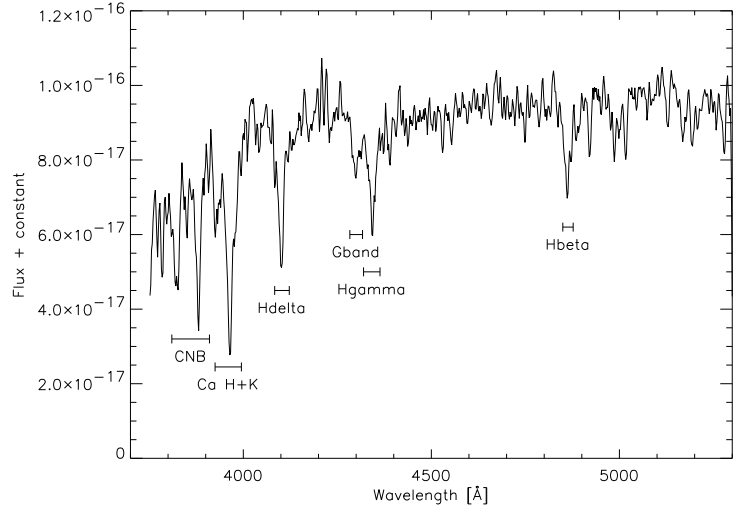,width=100mm}
  \caption{The LRIS blue side spectrum of H VIII, smoothed with a 3-pixel boxcar filter.}
\end{figure*}
\end{center} 

\begin{center}
\begin{figure*}
\psfig{file=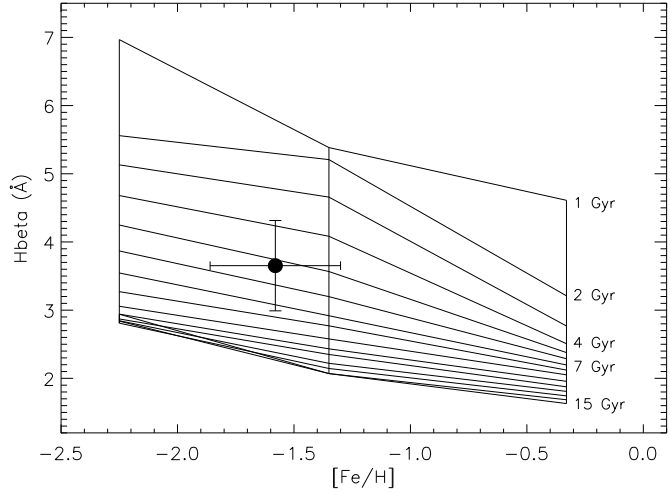,width=100mm}
  \caption{An [Fe/H] vs. H$\beta$ model grid. The isofers represent [Fe/H] = $-$2.25, $-$1.35, and $-$0.33.}
\end{figure*}
\end{center}

\begin{center}
\begin{figure*}
\psfig{file=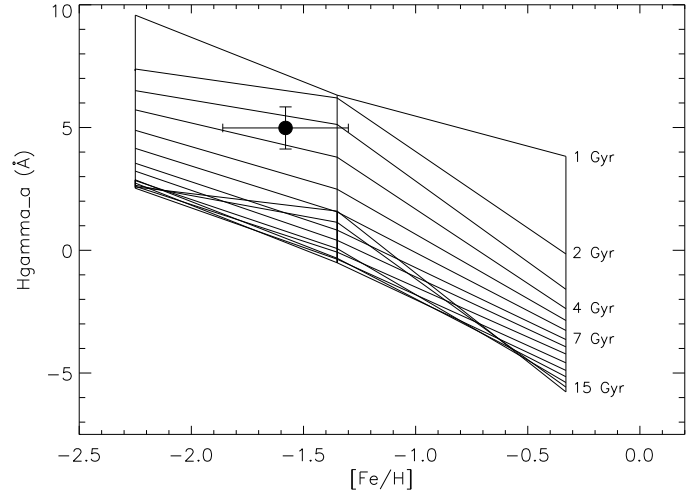,width=100mm}
  \caption{An [Fe/H] vs. H$\gamma_{A}$ model grid. The isofers represent [Fe/H] = $-$2.25, $-$1.35, and $-$0.33.}
\end{figure*}
\end{center}

\begin{table*}
 \centering
 \begin{minipage}{140mm}
  \caption{Metallicity Estimates from Brodie \& Huchra Indices}
  \begin{tabular}{ccccc}
  \hline
	$\Delta$       & G Band & CNB & CNR & [Fe/H] \\
 \hline
	$-1.27\pm0.37$ & $-1.72\pm0.47$ & $-1.08\pm0.44$ & $-2.17\pm0.48$ & $-1.58\pm0.28$ \\
\hline
\end{tabular}
\end{minipage}
\end{table*}

\begin{table*}
 \centering
 \begin{minipage}{140mm}
  \caption{Lick/IDS Indices}
  \begin{tabular}{ccccccc}
  \hline
	CN1 & CN2 & Ca4227 & G4300 & Fe5015 & Fe5782 & NaD \\
	(mag) & (mag) & (\AA) & (\AA) & (\AA) & (\AA) & (\AA) \\
 \hline
	$-0.18\pm0.02$ & $-0.15\pm0.03$ & $0.78\pm0.47$ & $2.05\pm0.92$ & $5.03\pm1.22$ & $1.01\pm0.32$ & $1.44\pm0.42$ \\
\hline
\end{tabular}
\end{minipage}
\end{table*}

\begin{table*}
 \centering
 \begin{minipage}{140mm}
  \caption{Balmer Line Indices}
  \begin{tabular}{cccccc}      
  \hline
	H$\beta$ & H$\gamma_{A}$ & H$\delta_{A}$ & H$\gamma_{F}$ & H$\delta_{F}$ & H$\alpha$\\
	(\AA)    & (\AA)          & (\AA)         & (\AA)         & (\AA)           & (\AA)\\
 \hline
	$4.19\pm0.66$ & $4.98\pm0.86$ & $8.97\pm0.94$ & $4.32\pm0.57$ & $6.07\pm0.68$ & $5.95\pm0.41$\\
\hline
\end{tabular}
\end{minipage}
\end{table*}

\end{document}